\documentclass[10pt,aps,prb]{revtex4}
\usepackage{amsmath,amssymb}
\begin{document}
\bibliographystyle{plain}

\title{Continous Temkin theory of interface}%

\author{Toni Ivas}%
\email{toni.ivas@mat.ethz.ch}
\affiliation{ETH Zurich, Switzerland}
\date{\today}
\begin{abstract}
We present differential equation for evolution of interface based on continuous approximation of Temkin's model. 
\end{abstract}

\maketitle


We start from a free energy functional of type~\cite{Schmitz}:
\begin{eqnarray}
\label{eq:1}
F[\phi]&=&\int a\nabla\phi\ln(a\nabla\phi)+\phi^{2}f^{S}+
(1-\phi)^2f^{L} \\ \nonumber
&+&2\phi(1-\phi)f^{LS}d^3x
\end{eqnarray}

Where $\phi$ is the order parameter equivalent to fraction of solid, 
and $1-\phi$  is the fraction of liquid in Temkin's model of 
solidification~\cite{Temkin}, we can also assume that $\phi$  represents fraction 
of phase A and $1-\phi$   fraction of phase B in an AB system. 
Consequently $f^S$, $f^L$ and $f^{LS}$ are free energies of pure solid, 
liquid and liquid-solid phases, respectively. In case of Temkin's model $a$  
is vector representing crystal lattice parameter, but in above functional 
we can generalize this parameter to represents a characteristic length of growth
not necessarily related to the lattice parameter.  
The lattice vector $a$ can be some function 
of the order parameter  which controls the evolution of the 
characteristic length and depend on additional space variables
including Euler angles but this further complicates our model. 
By doing the variational derivative of Eq.~(\ref{eq:1}) and defining 
$\alpha/4=f^{LS}-\frac{f^{L}+f^{S}}{2}$
and $\beta/2 = f^{LS}-f^{S}$  we have:
\begin{equation}
\label{eq:2}
\frac{1}{a\nabla\phi} \times
\left(a_{x}^2\partial_x^{2}+a_{y}^2\partial_y^{2}+a_{z}^2\partial_z^{2} 
+2a_{x}^2a_{y}^2\partial_{xy} 
+2a_{x}^2a_{z}^2\partial_{xz}+2a_{y}^2a_{z}^2\partial_{yz} \right)\phi-\alpha
\phi+\beta=\tau\frac{\partial\phi}{\partial t}
\end{equation}
Eq.~(\ref{eq:2}) describes the general case of anisotropy 
growth of interface in three dimensional
space. The parameters $\alpha$ and $\beta$ are defined as
$L/kT $ and $\Delta \mu / kT$, respectively. The Jackson parameter $\alpha$
is measure of surface energy and $\beta$ is difference in chemical potential
between two phases~\cite{Jackson}. 

We restrict our considerations only on isotropic case where: 
$a_x=a_y=a_z=a$ and one dimensional
case reducing Eq.(\ref{eq:2}) to: 
\begin{equation}
\label{eq:oned}
a\frac{\partial^{2}\phi}{\partial x^2}/\frac{\partial \phi}{\partial x}
-\alpha\phi + \beta = \tau \frac{\partial \phi}{\partial t}
\end{equation}
We will now give some solutions of previous evolution equation.
\section{Equilibrium solution}
The solution of differential equation~(\ref{eq:oned}) for equilibrium conditions 
$\beta=0$ and $\frac{\partial\phi}{\partial t}=0$ is: 
$\phi(x)=-\tanh(\alpha/2a(x-\delta))$ where 
$\delta$ is an arbitrary constant representing current position 
of the interface. The thickness of the interface is given by: 
\begin{equation}
\label{eq:equilib}
d=\frac{2a}{\alpha}
\end{equation}
This result is in agreement with Temkin's theory and
it's already given in \cite{HYN}. It predicts that 
interface thickness would depend on the parameter which 
Temkin defines as:
\begin{equation}
\label{eq:th}
\alpha=\frac{z\epsilon}{kT}=\frac{L}{kT}
\end{equation}
Where $z$ is number of nearest neighbors or 
coordination number and $\epsilon=\epsilon_{sl}-\frac{\epsilon_{ss}+\epsilon_{ll}}{2}$ 
is binding energy of the interface. Interesting consequence of 
relation~(\ref{eq:th}) is prediction of the thinner interface film for 
higher energy of binding when the thickness d is normalized 
with lattice constant a. This can be compared with experimental 
facts for different materials.

Using~Eq.(\ref{eq:th}) we can estimate the equilibrium thickness of the 
intergranular film for usual ceramic materials.
If we take approximate values for ceramics materials, 
the equilibrium temperature of about 2200 K and corresponding 
value for latent heat L = $1.5 *10^{-20}$ J/atom we get the thickness 
of intergranular film which is in agreement with the experimental findings 
and previous theoretical work of Bobeth et al.~\cite{Bobeth}. 

\section{Vector case}
Providing that we have slow growth of the crystal in x-direction 
we expect a slow change of $a_x$ due the small temperature gradient. 
Thus in the limiting case of the adiabatic change we can change 
$a_x$ with the constant value of $a$. In the general case $a_x$ 
will change its value with the time $a_x(t)$. We would like to describe 
the change of $a_x$ as a function of time and the simplest model 
we can imagine is a harmonic oscillator. Then we can define $a_x$ as:
\begin{equation}
a_{x}=a(1+\cos(\omega t))
\end{equation}
Where $\omega$ is the mean angular velocity of all atoms at the interface. 
Solution of our new differential equation in the case of vanishing chemical potential $\beta=0$ is:
\begin{eqnarray}
\label{eq:oscil}
\phi(x,t) &=& -\tanh(m(x-\delta)) \times \\ \nonumber 
& &\left[\frac{2am(\alpha^2+\omega^2+\alpha^2\cos(\omega t)+
\alpha\omega\sin(\omega t))}{\alpha(\alpha^2+\omega^2)}\right]
\end{eqnarray} 
Where $m\in~N $~Eq.(\ref{eq:oscil}) comes from the solution by separation of variables.
This is wave-like tangent hyperbolical solution contains two parts, first which is proportional to
thickness of the interface and second is harmonic oscillations 
of atoms at the interface. The harmonic oscillations of the diffuse interface will be 
govern by ratio $\alpha/\omega$. 
\section{Equilibrium solution in 2 D case}
In this section we map the solution of differential ~Eq.(\ref{eq:2}) 
in the case of cylindrical symmetry to solution of Burgers vector equation:
\begin{equation}
\label{eq:Burgers-vec}
\frac{\partial U}{\partial t}+U\nabla U=\nu\nabla^2 U
\end{equation}
This non-linear partial differential equation is described and solved 
in the detail in~\cite{Sachdev,Nerney}.  Main idea is to change 
the vector field U with the gradient of the scalar field 
$\psi$ imposing $\nabla\times U=0$ and $\nabla \psi=U$. 
Modifying Eq.~(\ref{eq:Burgers-vec}) we can simplify equation to:
\begin{equation}
\frac{\partial \psi}{\partial t}+(\frac{\nabla \psi}{2})^2-\nu\nabla^2 \psi=E(t)
\end{equation}
Where $E(t)$ is function of time only. Applying the same trick to our Eq.~(\ref{eq:2}) 
we get simplified version:
\begin{equation}
\label{eq:cyl}
a\nabla^2 u -\frac{\alpha}{2}(\nabla u) + \beta\nabla u 
- \frac{\partial \phi}{\partial t}\nabla u = 0
\end{equation}
Where $u = \nabla \phi$ and $\nabla \times u = 0$.
For case of the cylindrical symmetry and case where lattice 
vector $a$ is always parallel to normal of phase field, 
Eq.~(\ref{eq:cyl}) can be reduced to:
\begin{equation}
\label{eq:my-cyl}
a \frac{\frac{\partial}{\partial r}\left(\frac{1}{r}\frac{\partial}{\partial r}
(r\phi)\right)}{\frac{\partial \phi}{\partial r}}
- \alpha \phi + \beta = \tau \frac{\partial \phi}{\partial t}
\end{equation}
Equilibrium solution to Eq.~(\ref{eq:my-cyl}) is given as:
\begin{equation}
\phi(r) = -\frac{2a/\alpha}{(r-\delta)\ln(r-\delta)}
\end{equation}
As in the one dimensional case the thickness is characterized by $2a/\alpha$, but
it has logarithmic divergence to infinity.
\section{Dependence of the intergranular film thickness 
on the misfit angle and grain boundary angle}
The potential energy of grain boundary in Read-Shockley only depends on the 
dislocations in the crystal, produced by the mismatch of the crystal lattices. 
If the amorphous intergranular film is in the thermodynamic equilibrium 
with crystalline solid, the necessary condition for thermodynamic equilibrium 
is the minimum of the free energy leading to conclusion that amorphous 
intergranular film has a lower free energy than grain boundary between two crystalline solids. 
If we take that $\varphi$  is angle of grain boundary and $\theta$ 
is misfit angle, then from Read-Shockley theory of the 
grain boundary dislocations~\cite{Read} in the crystals we can write 
potential energy of grain boundary as:
\begin{equation}
    E=E_o\theta(A-\ln\theta)
\end{equation}
Necessary condition for existence of the amorphous intergranular layer is:
\begin{equation}
    E \geq E_{amorph}
\end{equation}
Where:
\begin{eqnarray*}
    &E_o&=Ga(\cos\varphi+\sin\varphi)/(4\pi(1-\sigma)) \\
    &E_{amorph}&-\mbox{energy of an amorphous-crystal interface} \\
    &A&-\mbox{function depending on $\varphi$} \\
    &\phi&-\mbox{misfit angle}
\end{eqnarray*}
We assume that energy of amorphous layer is exactly equal to grain boundary energy
of the Read-Schokley theory, which fulfil the necessary condition for 
thermodynamic equilibrium of amorphous film. In perfect solid the lattice vector $a$ is
oriented in one particular direction, while in the diffuse interface the lattice vector has
statistical distribution . This can be pictured as Ising XY-model 
where grain boundaries are regions in the plane where spins have different orientations
and regions of the crystal are perfectly oriented.

The probability of an orientation of a lattice vector in the space angle is
given by:
\begin{equation}
    dW_o=const\quad dK
\end{equation}
Or using the Boltzmann factor:
\begin{equation}\label{eq:boltz}
    dW=const\quad e^{E_o\theta(A-\ln\theta)/kT} \sin\varphi d\varphi
\end{equation}
\begin{eqnarray*}
\langle a_{x} \rangle&=&\frac{\int a_x dW}{\int dW} \\
&=&\frac{\int a \cos\varphi \exp[\frac{Ga(\cos\varphi + \sin\varphi)}{4\pi(1-\sigma)}\theta(A-\ln\theta)/kT]
\sin\varphi d\varphi}{\int \exp[\frac{Ga(\cos\varphi+\sin\varphi)}{4\pi(1-\sigma)}\theta(A-\ln\theta)/kT]
\sin\varphi d\varphi} \\
&=& a \sqrt 2 / 2 
\end{eqnarray*}
If we assume that $\cos\varphi \gg \sin\varphi$  which is true for small angles, the exponent in exponential function is simply $G a (\cos\varphi+\sin\varphi)\approx Ga\cos\varphi$, changing the integration variable   $x=a\cos\varphi G/ 4\pi(1-\sigma)\theta(A-\ln\theta)$ and collecting all parameters in $\lambda=\frac{G}{4\pi(1-\sigma)} \theta(A-\ln\theta)$  we have the following integral:
\begin{eqnarray}
\label{eq:stat_avg}
\left<a_x\right>&=&\frac{1}{\lambda} \frac{\int_0^{a\lambda} xe^x dx}{\int_0^{a\lambda} e^x dx} = \\ \nonumber
& & \frac{1}{\lambda}\frac{xe^x|_{0}^{a\lambda}-e^x|_{0}^{a\lambda}}{e^x|_{0}^{a\lambda}}
= \frac{1}{2\lambda}\frac{e^{a\lambda/2}(a\lambda-1)+e^{-a\lambda/2}}{\sinh(a\lambda/2)}
\end{eqnarray}
In most general cases $\lambda\approx 10^{26}$   is a big number for ceramic materials,so the value 
of $a_x$ can be calculated as $\lambda$ tends to infinity.
\begin{equation}
\lim_{\lambda->\infty}\frac{1}{2\lambda}
\frac{e^{a\lambda/2}(a\lambda-1)+e^{-a\lambda/2}}
{\sinh(a\lambda/2)}=a
\end{equation}
Thus we see that in limes for small misfit angles and the condition 
that  $\cos\varphi \gg \sin\varphi$ we have uniform thickness 
of the intergranular films. This is in good agreement with the experiments.
Lets calculate the special case where $\lambda=0$, this the
 case when angle of misfit is $\theta=0$.
\begin{equation}
\lim_{\lambda->0}\frac{1}{2\lambda}\frac{e^{a\lambda/2}(a\lambda-1)+e^{-a\lambda/2}}{\sinh(a\lambda/2)}=a/2
\end{equation}
Previous calculation can be used together with~Eq.(\ref{eq:equilib}) 
to calculate the thickness $\langle d \rangle=\frac{2 \langle a_{x} \rangle}{\alpha}$ of intergranular film
in polycrystalline material.
From this consideration we conclude that average 
thickness over wide range of crystallographic 
orientations is almost constant and it reduces to 
half this value only in the case where the misfit angle is zero.

\section{Kinetics of the intergranular film}
The kinetics of interface in one dimensional case is allready given by Mori et al.~\cite{Mori} 
Solution of Eq.~(\ref{eq:2}) is given as:
\begin{equation}
    \label{eq:vel_sol}
    \tanh(v)=\frac{v}{(\beta/\alpha) v +1}
\end{equation}
Where $v=V\tau/a$ is non-dimensional velocity of the interface and $V$ is
real velocity of the interface.
Velocity diverges at point where $\beta\rightarrow\alpha$ 
this is non-physical solution
but the validity of Time-Dependent-Ginsburg-Landau equation 
(TDGL) is questionable in the range of very high 
velocity of the interface. 
We can give several limiting cases for Eq.~(\ref{eq:vel_sol}):
\begin{eqnarray*}
    V \gg a/\tau \quad (\beta \nearrow \alpha): \\
    v \simeq -1/2 \ln\frac{1-(\alpha/\beta)}{2} \\
    -V \gg a/\tau \quad (\beta \searrow -\alpha): \\
    v \simeq 1/2 \ln\frac{1+(\alpha/\beta)}{2}
\end{eqnarray*}

We conclude with application of evolution equation 
to phase separation in polymer mixtures. 
This is usually described with Flory-Huggins free energy given as:
\begin{equation}
 \frac{F}{kT} = \phi_{A}\ln\phi_{A}+\phi_{B}\ln\phi_{B} 
  + N \chi\phi_{A}\phi_{B} 
\end{equation}
Where N is degree of polymerization, $\chi$ interaction parameter 
per monomer (Flory-Huggins parameter),  $\phi_A$ and $\phi_B$ is probability of site in space 
is occupied with molecule of A or molecule of B species, respectively. 
Lets take a characteristic length of polymer chain to be is given by 
$w=a\sqrt{N}$ where a is size of monomer and N characteristic loop size of 
random walk of polymer A protruding in space of polymer B. 
Then total energy coming from interaction between A and B segments of
polymer is $L=\chi NkT$ so $\alpha=\frac{L}{kT}=\chi N$. 
If we put this result inside Eq.~(\ref{eq:equilib}) we have:
\begin{equation}
d=\frac{2w}{\alpha}=\frac{2a\sqrt{N_{loop}}}{\chi N_{loop}}
=\frac{2a}{\chi\sqrt{N_{loop}}}
\end{equation}
\\
For equilibrium the interaction energy of monomer will be 
of order $kT$  thus: $\chi N_{loop}\approx 1$.
Giving finally in equilibrium case:
\begin{equation}
d \approx \frac{2a}{\sqrt{\chi}}
\end{equation}
Which is of course not a surprise as Temkin's model is equivalent
to Flory-Huggins theory of polymer mixtures.
\bibliography{interface_film_theory}
\section{Appendix}
\begin{equation}
\frac{\delta f}{\delta \phi} =
\frac{\partial f}{\partial \phi}-\sum\frac{\partial}{\partial x_i}
\left (\frac{\partial f}{\frac{\partial \phi}{\partial x_i}} \right )
\end{equation}
\begin{equation}
\frac{\partial f}{\partial \phi_x} =
\frac{\partial}{\partial \phi_x} \left ((a_x \phi_x + a_y \phi_y +
a_z \phi_z)  \ln (a_x \phi_x + a_y \phi_y + a_z \phi_z)\right)
\end{equation}
Then
\begin{eqnarray}
& & \frac{\partial}{\partial x}\left  ( \frac{\partial f}{\partial \phi_x}\right ) \\ \nonumber
& = & \frac{\partial}{\partial x}\left (a_x \ln (a_x \phi_x + a_y \phi_y + a_z
\phi_z ) + a_{x}^{2}\phi_x \frac{1}{a\nabla \phi}+
a_{y}\phi_y \frac{1}{a\nabla \phi}a_x+a_{z}\phi_z \frac{1}{a\nabla \phi}a_x
\right ) \\ \nonumber
& = & \frac{a_x}{a\nabla \phi}(a_x \phi_{xx} + a_y \phi_{yx} + a_z \phi_{zx}) +
a_{x}^{2}\phi_{xx}\frac{1}{a\nabla\phi} - a_{x}^2\phi_x\frac{1}{(a\nabla\phi)^2}
(a_x\phi_{xx}+a_y\phi_{yx}+a_z\phi_{zx}) \\ \nonumber
& + & a_y\phi_{yx}\frac{1}{a\nabla\phi}a_x-a_y\phi_y\frac{a_x}{(a\nabla\phi)^2}
(a_x \phi_{xx} + a_y \phi_{yx}+a_z\phi_{zx}) + a_z\phi_{zx}\frac{1}{a\nabla\phi}a_x \\ \nonumber
& - & a_z\phi_z\frac{a_x}{(a\nabla\phi)^2} (a_x\phi_{xx} + a_y\phi_{yx}+a_z\phi_{zx}) \\ \nonumber
& = & \frac{a_x}{a\nabla\phi}(a_x\phi_{xx}+a_y\phi_{yx}+a_z\phi_{zx})-
\frac{a_x}{(a\nabla\phi)^2}a\nabla\phi(a_x\phi_{xx}+a_y\phi_{yx}+a_z\phi_{zx}) \\ \nonumber
& + & \frac{a_x}{a\nabla\phi}(a_x\phi_{xx}+a_y\phi_{yx}+a_z\phi_{zx}) \\ \nonumber
& = & \frac{a_x}{a\nabla\phi}(a_x\phi_{xx}+a_y\phi_{yx}+a_z\phi_{zx})
\end{eqnarray}
Similarly, for components y and z we have :
\begin{eqnarray}
\frac{\partial}{\partial y}\left (\frac{\partial f}{\partial \phi_y} \right )= 
\frac{a_y}{a\nabla\phi}\left (a_x\phi_{xy}+a_y\phi_{yy}+a_z\phi_{zy}\right ) \\
\frac{\partial}{\partial z}\left (\frac{\partial f}{\partial \phi_z} \right )= 
\frac{a_z}{a\nabla\phi}\left (a_x\phi_{xz}+a_y\phi_{yz}+a_z\phi_{zz}\right )
\end{eqnarray}
Then the sum of these terms is:
\begin{eqnarray}
& &\frac{\partial}{\partial x}\left (\frac{\partial f}{\partial \phi_x}\right )+
\frac{\partial}{\partial y}\left (\frac{\partial f}{\partial \phi_y}\right )+
\frac{\partial}{\partial z}\left (\frac{\partial f}{\partial \phi_z}\right ) = \\ \nonumber
& &\frac{1}{a\nabla \phi}(a_{x}^2\phi_{xx}+a_{y}^2\phi_{yy}+a_{z}^2\phi_{zz}+
2a_x a_y \phi_{xy}+2a_x a_z\phi_{xz}+2a_y a_z\phi_{yz}) \\ \nonumber
& = & \frac{(a\nabla)^2\phi}{a\nabla\phi}
\end{eqnarray}
\subsection{One-dimensional solution to evolution equation}
We start from partial differential equation:
\begin{equation}
a\frac{\partial^2\phi}{\partial x^2}/\frac{\partial \phi}{\partial x} - \alpha\phi
+\beta = \tau\frac{\partial \phi}{\partial t}
\end{equation}
with change of variables: 
$u(\phi)=a\frac{\partial\phi}{\partial x}$ we have:
\begin{equation}
\frac{\partial u(\phi)}{\partial \phi} = \frac{\partial}{\partial \phi}\left (a\frac{\partial\phi}
{\partial x}\right)=a\frac{\partial^2\phi}{\partial x^2}\frac{\partial x}{\partial \phi}=
a\frac{\partial^2\phi}{\partial x^2}
/\frac{\partial\phi}{\partial x}
\end{equation}
$\frac{\partial \phi}{\partial t} = -V\frac{\partial \phi}{\partial x}$
and we transform the equation in :
\begin{equation}
\frac{\partial u}{\partial \phi}+v u -\alpha\phi+\beta=0
\end{equation}
\begin{equation}
u(\phi)=Ce^{-v\phi}-\frac{\alpha}{v}(\frac{1}{v}-\phi)-\frac{\beta}{v}
\end{equation}
where\\
\begin{equation}
C=\sqrt{\left (\frac{\alpha}{v^2}+\frac{\beta}{v} \right)^2-
\left(\frac{\alpha}{v}\right)^2}
\end{equation}
Final solution is:
$\tanh v = \frac{v}{(\beta/\alpha)v + 1}$ which for $\beta=0$ gives $\tanh v = -v$.
\subsection{One-dimensional solution of evolution equation with harmonic oscillations}
$\frac{\partial \phi}{\partial t}=\beta-\alpha\phi+a(1+\cos(\omega t))
\frac{\partial^2\phi}{\partial x^2}/\frac{\partial \phi}{\partial x}$
\\
Lets solve it by separation of variables:
\begin{equation}
\phi(x,t)=X(x)T(t)
\end{equation}
Where X(x) is only function of x and T(t) is only function of t.
Then, derivatives are:
\begin{eqnarray}
\frac{\partial^2 \phi}{\partial x^2} & = & T(t)\frac{\partial^2 X}{\partial x^2} \\
\frac{\partial \phi}{\partial x} & = &T(t) \frac{\partial X}{\partial x} \\
\frac{\partial \phi}{\partial t} & = &\frac{\partial T}{\partial t} X(x)
\end{eqnarray}
Substituting these derivatives in differential equation:
\begin{equation}
a(1+\cos(\omega t))\frac{\partial^2 X}{\partial x^2}/\frac{\partial X}{\partial x}
-\alpha XT +\beta=X\frac{\partial T}{\partial t}
\end{equation}
Dividing this differential equation with X we get:
\begin{equation}
\frac{1}{X}a(1+cos(\omega t))\frac{\partial^2 X}{\partial x^2}/\frac{\partial X}{\partial x}
-\alpha T + \beta / X = \frac{\partial T}{\partial t}
\end{equation}
If search for solution of where there is no driving force for transition so we have
$\beta = 0$:
\begin{equation}
\frac{1}{X}\frac{\partial^2 X}{\partial x}/\frac{\partial X}{\partial x}
-(\alpha T+\frac{\partial T}{\partial t})\frac{1}{a(1+\cos(\omega t)} = 0
\end{equation}
This equation is separable, and we get two ordinary differential equations:
\begin{equation}
\frac{d^2 X}{dx^2}-2mX\frac{dX}{dx} = 0
\end{equation}
Where m is integer:
\begin{equation}
\frac{dT}{dt}+\alpha T-2ma(cos(\omega t)+1)=0
\end{equation}
The solution of second differential equation is simply:
\begin{equation}
T(t) = C \exp(-\alpha t) + \frac{2 am}{[\alpha^2+\omega^2+\alpha^2\cos(\omega t)
+\alpha\omega \sin(\omega t) ]}{\alpha(\alpha^2 + \omega^2)}
\end{equation}
The solution to first differential equation can be found by putting solution
X(x) back in the equation:
\begin{equation}
X(x) = -\tanh(m(x-\delta))
\end{equation}
So final solution is:
\begin{equation}
\phi(x,t)= -\tanh(m(x-\delta))\left[\frac{2am(\alpha^2 + \omega^2
+ \alpha^2\cos(\omega t)+\alpha\omega\sin(\omega t))}{\alpha
(\alpha^2+\omega^2)}\right ]
\end{equation}
\subsection{Calculation of Read-Schokley integral}
Calculation of the integral:
\begin{eqnarray}
& & \langle a_x\rangle= \frac{\int a_x dW}{\int dW}  =  \\ \nonumber 
& & \frac{\int \cos\phi \exp\left [ 
Ga(\cos\phi+\sin\phi)/(4\pi(1-\sigma))\theta (A-\ln \theta)/kT 
\right ] \sin\phi d\phi}
{\int \exp\left [Ga(\cos\phi+\sin\phi)/(4\pi(1-\sigma))\theta(A-\ln\theta)/kT\right]
\sin\phi d\phi}
\end{eqnarray}
We will evaluate this integral with method given in Massida. All constant factors
can be extracted from previous integral and then we need to solve general 
integral of type:
\begin{equation}
\int_{0}^{2\pi}e^{in\phi}\exp(\cos\phi+\sin\phi)d\phi
\end{equation}
Using relation between exponential and Bessel function:
\begin{eqnarray}
\exp(\frac{z}{2}(t+\frac{1}{t})) &=& \sum_{k=-\infty}^{k=\infty}t^k I_k(z) \\ 
e^{\cos\phi} = \sum_k t^k I_k(1) &=& \sum_{k}e^{ik\phi}I_k(1) \\
e^{\sin\phi} & = & \sum_h(-i)^h e^{ih\phi} I_h(1)
\end{eqnarray}
Thus integral can be written as:
\begin{equation}
\int_{0}^{2\pi}e^{in\phi}\sum_h(-i)^h e^{ih\phi}I_h(1)\sum_k e^{ik\phi}I_k(1)d\theta
\end{equation}
Using $I_{-m}(x)=I_{m}(x)$ we can transform integral to $2\pi\sum_k I_k(1)I_{k+1}(-i)^k$
using addition theorem for Bessel functions:
\begin{equation}
\sum_n (-1)^{n}e^{in\phi}I_n(z)I_{n+v}(Z)=I_v(\omega)e^{iv\psi}
\end{equation}
Where $\omega=(z^2+Z^2-2zZcos\phi)^{1/2}$ and $\psi$ is defined by $Z-z\cos\phi=
\omega\cos\psi$ and $z\sin\phi=\omega\sin\psi$. The addition theorem can then be
applied to k summation in sum and we get solution as:
\begin{equation}
2\pi\sum_k I_k(1)I_{k+1}(1)(-i)^k = I_1(\sqrt{2(1-cos\phi)})
e^{i \arccos(\sqrt{2(1-cos\phi)}/2)}
\end{equation}
Final solution to integral is:
\begin{equation}
\int_{0}^{2\pi}\cos(x)\exp\left[\cos(x)+\sin(x)\right]dx = 2\pi I_1(\sqrt{2})
\frac{\sqrt{2}}{2}
\end{equation}
Similarly for:
\begin{equation}
1/2 \int_{0}^{2\pi} \sin(2x) \exp\left[\cos(x)+\sin(x)\right]dx=I_v(\omega)e^{iv\phi}
= I_2(\sqrt{2})2\pi
\end{equation}
\subsection{Two-dimensional solution of evolution equation}
We start from partial differential equation:
\begin{equation}
a\frac{\frac{\partial }{\partial r}\left (\frac{1}{r}\frac{\partial}{\partial r}
(r\phi)\right )}{\frac{\partial \phi}{\partial r}}-\alpha\phi+\beta=
\frac{\partial \phi}{\partial t}\tau
\end{equation}
Using transformation from Nerney et al.~\cite{Nerney} we can write previous equation:
\begin{equation}
\frac{d\theta}{d\phi}-\alpha \phi+\beta= -v\theta(\phi)
\end{equation}
Where $v=\frac{V\tau}{a}$ non-dimensional velocity, solution to eqution is:
\begin{equation}
\theta=C e^{v\theta}+\frac{\alpha}{v}\phi-\beta/v-\alpha/v^2
\end{equation}
Where $C=\sqrt{\left(\frac{\alpha}{v^2}+\frac{\beta}{v}\right)
-\left(\frac{\alpha}{v}\right)^2}$.
For equilibrum conditions is equation reduced to:
\begin{equation}
a\frac{\partial}{\partial r}\left (\frac{1}{r}\frac{\partial}{\partial r}
(r\phi)\right )-\frac{\partial \phi}{\partial r}\alpha \phi = 0
\end{equation}
Solution of partial differential equation is:
\begin{equation}
\frac{1}{r}\frac{\partial}{\partial r}(r\phi)=\frac{C}{a}e^{\phi\alpha}
\end{equation}
Determination of constant of intergration C:
\begin{equation}
\phi(r) = 
  \begin{cases}
  1\; \text{when $r = 0$}, \\
  -1\; \text{when $r =\infty$} 
  \end{cases}
\end{equation}
Giving $C=0$ and finally the solution to differential equation:
\begin{equation}
\phi(r)=-\frac{2a/\alpha}{r\ln r}
\end{equation}

\end{document}